\documentstyle[preprint,aps,epsfig]{revtex}

\title{Coulomb effects in nucleon-deuteron polarization-transfer coefficients} 
\author{A. Kievsky, S. Rosati and M. Viviani}
\address{Istituto Nazionale di Fisica Nucleare, Via Buonarroti 2, 56100 Pisa, 
Italy} 
\address{Dipartimento di Fisica, Universita' di Pisa, Via Buonarroti 2,
56100 Pisa, Italy}

\begin{document}

\maketitle

\begin{abstract}
Coulomb effects in the neutron-deuteron and proton-deuteron 
polarization-transfer 
coefficients $K_y^{y'}$, $K_z^{x'}$, $K_y^{x'x'-y'y'}$ and $K_y^{z'z'}$
are studied at energies above the deuteron breakup threshold. 
Theoretical predictions for these observables are evaluated
in the framework of the Kohn Variational Principle
using correlated basis functions to expand the three-nucleon scattering
wave function. The two-nucleon Argonne $v_{18}$ and the
three-nucleon Urbana IX potentials are considered. In the 
proton-deuteron case, the Coulomb interaction between the two protons 
is included explicitly and the results are compared to 
the experimental data available at $E_{lab}=10,19,22.7$ MeV. In the
neutron-deuteron case, a comparison to a recent measurement of $K_y^{y'}$ 
by Hempen {\sl et al.} at $E_{lab}=19$ MeV evidences
a contribution of the calculated Coulomb effects opposite
to those extracted from the experiment. 
\end{abstract}

\newpage

The three-nucleon (3N) system is an excellent testing ground for
the nuclear interaction. The last generation NN interactions
can be used to calculate 3N bound and scattering states and
important conclusions about
the capability of those interactions to reproduce the 3N dynamics
can be derived from a comparison to the experimental data. 
In the 3N system the potential energy consists of a 
sum of the pairwise NN interaction and a term including a pure
three-nucleon interaction (TNI). The TNI term is not very well known and,
in general, its strength is fixed so as to reproduce the experimental
$A=3$ binding energy. Due to recent advances
in the solution of the 3N continuum,
the possibility of using the 3N scattering data
to improve the TNI is at present feasible.
Because of this, a correct treatment 
of the Coulomb interaction in the description of the proton-deuteron reaction
is required, which has been a difficult problem until recently.
However, the 3N continuum is largely dominated by the NN interaction, so
the specific sensitivity of particular observables to the TNI is of
interest.

Here we focus attention on the polarization-transfer
coefficients $K_y^{y'}$, $K_z^{x'}$ and  $K_y^{x'x'-y'y'}$, $K_y^{z'z'}$
in the three-nucleon system.
These coefficients are sensitive to the tensor force and their study
from an experimental and theoretical point of view 
provides information about parts of the nuclear interaction not
completely under control. For example, in the 2N system the mixing 
parameter $\epsilon_1$,
which is directly related to the tensor force, has been studied very
recently in a double polarized neutron-proton experiment at low 
energy~\cite{tornow}. The results appear to be in disagreement with the
predictions of the modern NN interactions. On the other hand, the
calculations of $\epsilon_1$ and several others 2N phase-shift and mixing 
parameters using these new 
interactions, which describe the 2N data with $\chi^2/N$
$\approx 1$, show non-negligible differences between the 
different models. These differences are related possibly to an incomplete 
database or to a low sensitivity of some parameters to a large number
of observables. To pin down the non-central parts of the NN interaction 
necessitates measurements using polarized beams or targets which, 
in general, are difficult to perform. Though the picture of the 2N system
is still open to for improvements, it is natural to extend the study
of the nuclear interaction to the 3N system.

The polarization-transfer
coefficients $K_y^{y'}$, $K_z^{x'}$ and  $K_y^{x'x'-y'y'}$, $K_y^{z'z'}$
have been measured very recently 
in the elastic reaction $D(\vec{p},\vec{p})d$ 
and $D(\vec{p},\vec{d})p$ at $E_{lab}=19$ MeV, respectively~\cite{sydow}.
Furthermore, the coefficient $K_y^{y'}$ has been measured at the same
energy for the neutron-deuteron reaction~\cite{hempen} allowing
interesting comparisons to calculations with or without the Coulomb
interaction. In the mentioned works, a number of calculations for these 
observables have been performed by the Bochum group solving the Faddeev
equations in momentum space. Different modern NN interactions has been
considered as well as the Tucson-Melbourne TNI. 
The main conclusions from refs.~\cite{sydow,hempen} are: $i$) the
nucleon coefficients $K_y^{y'}$, $K_z^{x'}$ show a scaling behavior
with the triton binding energy whereas the deuteron coefficients
$K_y^{x'x'-y'y'}$, $K_y^{z'z'}$ do not, and $ii$) the nucleon coefficient
$K_y^{y'}$ shows appreciable Coulomb effects in its minimum at
the center of mass angle $\theta_{c.m.}\approx 110^\circ$. 
A definitive conclusion about
the capability of the theory to describe these coefficients cannot
be extracted from ref.~\cite{sydow} since the Coulomb interaction was 
neglected in the calculations. On the other hand, in the case of
$K_y^{y'}$ it was possible to make a direct comparison to the n-d 
data of ref.~\cite{hempen}. A spreading of the predictions of the pure 
NN forces at the minimum of $K_y^{y'}$ has been observed. 
This spreading has been
related to differences in the deuteron $D$--state probability and
the triton binding energy given by the different NN models.
However, when the TNI was included, and its cutoff dependence fixed to 
reproduce the experimental value of the triton binding energy, the 
predictions did not show such a spreading any more.
All calculations underestimated the minimum by about $10$\%.

The main results of the present communication are given in figs.1-3.
In figs.1 and 2 the polarization-transfer coefficients calculated
for a few potential models at $E_{lab}=19$ MeV are shown. In fig.3 the
results for $K_y^{y'}$ are given at other two energies, $E_{lab}=10$
and $22.7$ MeV.
The calculations have been made using the Kohn
variational principle (KVP) with an expansion of the scattering wave
function in terms of the pair correlated hyperspherical harmonic 
basis~\cite{phh}. The use of the KVP to describe proton-deuteron
scattering at energies above the deuteron
breakup has been discussed by the authors in ref.~\cite{VKR00}. 
Details of the method are given in ref.~\cite{KVR01} together with
results of the nucleon-deuteron cross section and vector and
tensor analyzing powers up to $E_{lab}=30$ MeV.

In fig.1 the theoretical predictions for $K_y^{y'}$ and $K_z^{x'}$ are 
compared to the experimental data of refs.~\cite{sydow,hempen}. 
Calculations for proton-deuteron scattering
have been made using the Argonne $v_{18}$ (AV18) 
potential~\cite{av18} (solid line) and also with the inclusion of 
the Urbana (UR)~\cite{urbana} TNI (dotted line). 
The potential model AV18+UR has 
the property of reproducing the $^3$He binding energy~\cite{KVR95}. 
A neutron-deuteron calculation using AV18 is also shown 
(long-dashed line). Results are given in fig.2 for the coefficients 
$K_y^{x'x'-y'y'}$ and $K_y^{z'z'}$ and in fig.3
for the coefficient $K_y^{y'}$ at $E_{lab}=10$ and $22.7$ MeV.
Experimental data are from refs~\cite{sydow,sperisen,clajus}.

Let us discuss first the results for the nucleon coefficients
$K_y^{y'}$ and $K_z^{x'}$ at $E_{lab}=19$ MeV. 
The p-d AV18+UR and the n-d AV18 curves
are very close to each other showing that Coulomb and TNI effects 
are of the same size. Moreover, the AV18+UR calculations for p-d
scattering reproduce the p-d data reasonably well. A comparison of the
AV18 p-d and n-d curves shows that the Coulomb interaction tends to
increase the absolute values of 
$K_z^{x'}$ and $K_y^{y'}$ at $\theta_{c.m.}\approx 110^\circ$
in the maximum and the minimum, respectively. Conversely,
the n-d $K_y^{y'}$ datum at the minimum (shown by a square in fig.1)
is above the corresponding p-d datum.
In other words, the inclusion of the Coulomb interaction in the
calculations increases
$K_y^{y'}(110^\circ)$, contrary to the experimental
observation.
The fact that the n-d curve describes the n-d $K_y^{y'}(110^\circ)$ datum 
should be taken
with caution since when the Tucson-Melbourne TNI is included
the theoretical n-d predictions shift below the p-d data, as has been
analyzed in ref.~\cite{hempen}. 
This conflict between theory and experiment at relatively low energy 
is of interest. However a larger number of data points seems to be
necessary before formulating a final conclusion.

The results for the deuteron coefficients $K_y^{x'x'-y'y'}$ and $K_y^{z'z'}$
given in fig.2 are less sensitive to Coulomb and TNI effects. 
The p-d AV18 curve seems to underestimate the maximum of $K_y^{x'x'-y'y'}$
as well as to overestimate the minimum of $K_y^{z'z'}$ both
at $\theta_{c.m.}\approx 130^\circ$. 
The p-d AV18+UR curve slightly improves the description, but
larger TNI effects seem to be necessary to correctly
describe these peaks. Certainly smaller error bars would allow a more
conclusive analysis.

In fig.3 the energy dependence of $K_y^{y'}$ can be analyzed. At 
$E_{lab}=10$ MeV the three curves are close to each other and describe
reasonably well the experimental points of ref.~\cite{sperisen}. The
minimum observed at higher energies is very shallow here and Coulomb
and TNI effects are small. Conversely, at $E_{lab}=22.7$ MeV
we observe a well pronounced minimum in which Coulomb and TNI effects
are even larger than at $E_{lab}=19$ MeV. Again the 
AV18+UR curve gives the better description of the data of ref.~\cite{clajus}.

In conclusion, there are appreciable Coulomb and TNI effects in the 
polarization transfer coefficients. The two contributions are
of the same order of magnitude, the same situation being observed in
the calculation of the three-nucleon binding energy. In fact, the triton
binding energy calculated with any of the new local NN interactions
is $B_t\approx 7.6$ MeV~\cite{friar}. When the Coulomb potential
and TNI terms are taken into account, with the TNI strength fixed 
to reproduce the experimental binding, the value 
$B(^3{\rm He})\approx 7.7$ MeV is obtained. So, the scaling
property found in the coefficients $K_y^{y'}$ and $K_z^{x'}$ explains
why the p-d AV18+UR and the n-d AV18 are close to each other. 
On the other hand,
the explanation of the opposite value of
the calculated Coulomb effects and those extracted from the 
experimental data is not clear at the moment.
New measurements of the nucleon-deuteron polarization transfer coefficients
could be useful for a more extensive analysis of the discrepancies observed.

\newpage

Figure captions

\bigskip

Fig.1. The polarization-transfer coefficients $K_z^{x'}$ and 
       $K_y^{y'}$ at $E_{lab}=19$ MeV. 
       Calculations are shown for p-d 
       scattering using the AV18 (solid line) and AV18+UR (dotted line)
       potentials and for n-d scattering using the AV18 potential
       (long-dashed line). 
       Experimental
       points for p-d scattering (circles) and for n-d scattering
       (squares) are from refs.~\cite{sydow,hempen} respectively.

Fig.2. The polarization-transfer coefficients $K_y^{x'x'-y'y'}$ and 
       $K_y^{z'z'}$ at $E_{lab}=19$ MeV. 
       Calculations are shown for p-d 
       scattering using the AV18 (solid line) and AV18+UR (dotted line)
       potentials and for n-d scattering using the AV18 potential
       (long-dashed line). 
       Experimental points for p-d 
       scattering (circles) are from ref.~\cite{sydow}.

Fig.3. The polarization-transfer coefficient
       $K_y^{y'}$ at $E_{lab}=10$ and $22.7$ MeV. 
       Calculations are shown for p-d 
       scattering using the AV18 (solid line) and AV18+UR (dotted line)
       potentials and for n-d scattering using the AV18 potential
       (long-dashed line). 
       Experimental points for p-d 
       scattering (circles) are from ref.~\cite{sperisen} ($10$ MeV)
       and from ref.~\cite{clajus} ($22.7$ MeV).

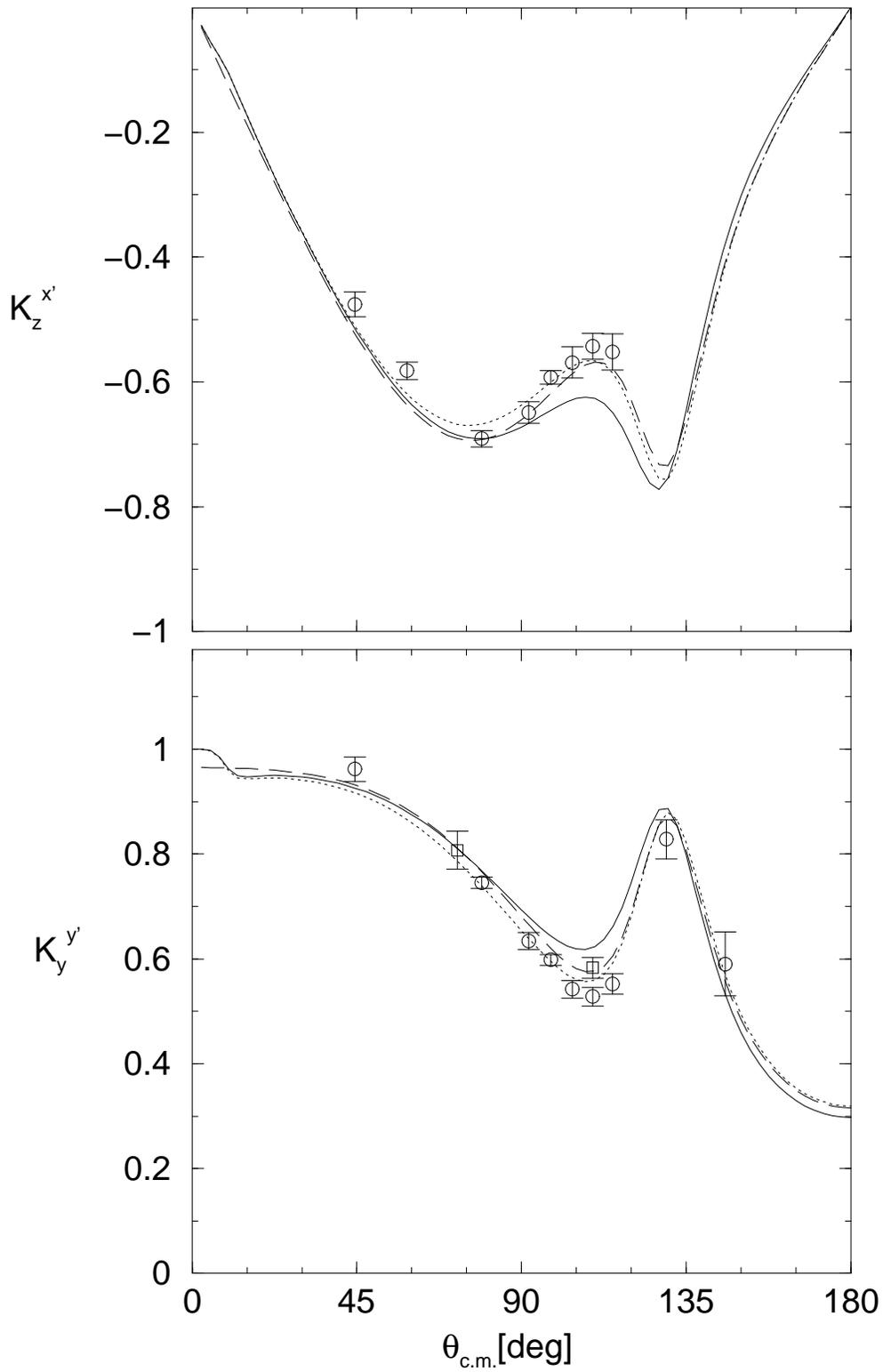
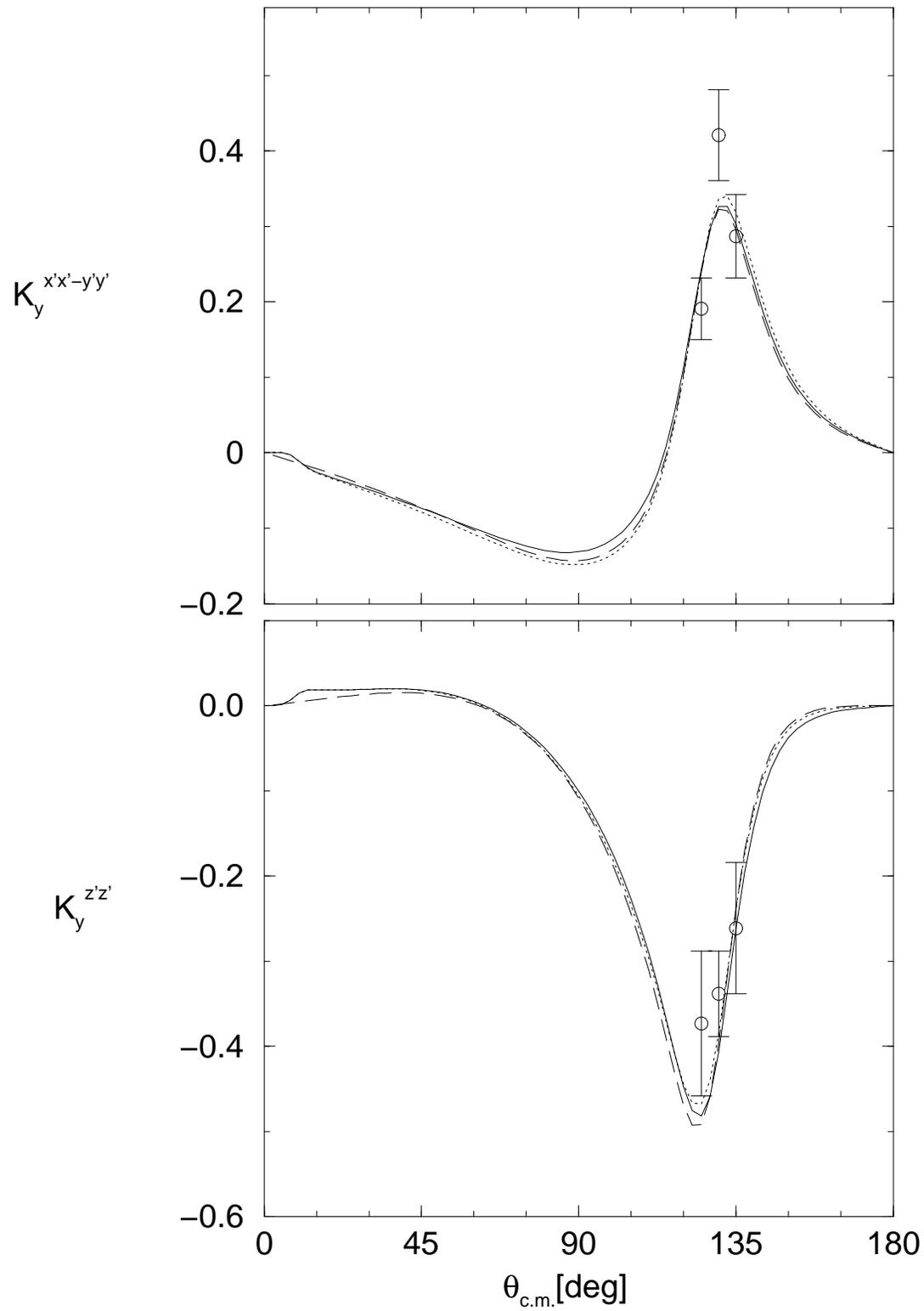
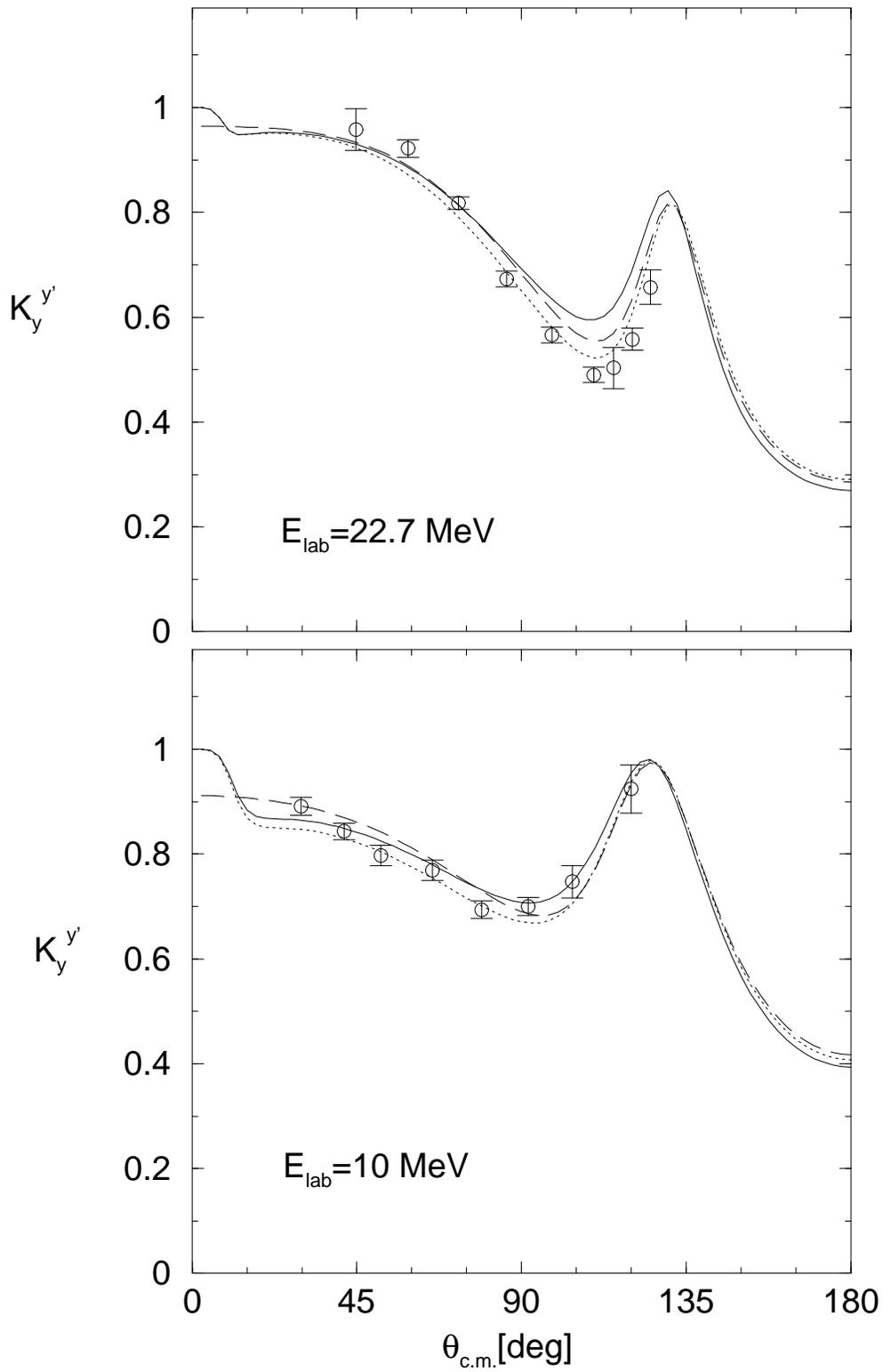
\end{document}